\documentclass[conference]{IEEEtran}
\IEEEoverridecommandlockouts
\usepackage{cite}
\usepackage{amsmath,amssymb,amsfonts,amsthm}
\usepackage{bm}
\usepackage{array}
\usepackage{mdwmath}
\usepackage{mdwtab}
\usepackage{eqparbox}
\usepackage[overload]{empheq} 
\usepackage{algorithmic}
\usepackage{graphicx}
\usepackage{textcomp}
\usepackage{xcolor}
\usepackage[most]{tcolorbox}
\usepackage{acronym}
\usepackage{tikz} \usetikzlibrary[topaths] 
\usetikzlibrary{arrows.meta}
\usetikzlibrary{positioning}
\usetikzlibrary{bending}
\usepackage{pgfplots}
\pgfplotsset{compat=1.18}
\usepgfplotslibrary{polar}
\usepackage[caption=false,font=footnotesize]{subfig}
\usepackage[export]{adjustbox}
\usepackage{booktabs,makecell,tabularx}
\usepackage{multirow}
\usepackage{geometry}
\usepackage{afterpage}
\usepackage[bookmarks=true,hidelinks,pdfpagelabels=true]{hyperref}
\usepackage{microtype}
\def\BibTeX{{\rm B\kern-.05em{\sc i\kern-.025em b}\kern-.08em
    T\kern-.1667em\lower.7ex\hbox{E}\kern-.125emX}}
    
\definecolor{lightgrey}{rgb}{0.85, 0.85, 0.85}
\definecolor{morelightgrey}{rgb}{0.95, 0.95, 0.95}
\definecolor{amber}{rgb}{1.0, 0.75, 0.0}
\definecolor{airforceblue}{rgb}{0.36, 0.54, 0.66}
\definecolor{aquamarine}{rgb}{0.5, 1.0, 0.83}
\definecolor{azure}{rgb}{0.0, 0.5, 1.0}
\definecolor{cadmiumorange}{rgb}{0.93, 0.53, 0.18}
\definecolor{babypink}{rgb}{0.96, 0.76, 0.76}
\definecolor{bubblegum}{rgb}{0.99, 0.76, 0.8}
\definecolor{alizarin}{rgb}{0.82, 0.1, 0.26}
\definecolor{asparagus}{rgb}{0.53, 0.66, 0.42}
\definecolor{cadetgrey}{rgb}{0.57, 0.64, 0.69}
\definecolor{wildstrawberry}{rgb}{1.0, 0.26, 0.64}
\definecolor{green_python}{HTML}{2ca02c}
\definecolor{orange_python}{HTML}{ff7f0e}
\definecolor{aliceblue}{rgb}{0.94, 0.97, 1.0}
\definecolor{cherryblossompink}{rgb}{1.0, 0.72, 0.77}
\definecolor{cerulean}{rgb}{0.0, 0.48, 0.65}
\definecolor{darkcerulean}{rgb}{0.03, 0.27, 0.49}
\definecolor{bluegray}{rgb}{0.4, 0.6, 0.8}
\definecolor{apricot}{rgb}{0.98, 0.81, 0.69}
\definecolor{desertsand}{rgb}{0.93, 0.79, 0.69}
\definecolor{coloreaggiunta}{HTML}{F6D7DF}
\definecolor{applegreen}{rgb}{0.55, 0.71, 0.0}

\begin{document}
\newgeometry{left=54pt,right=54pt,top=72pt,bottom=54pt}%

  \acrodef{AsD}[AsD]{Assisted Driving}
  \acrodef{ASPI}[ASPI]{Autostrade Per l'Italia}
  \acrodef{AuD}[AuD]{Automated Driving}
  \acrodef{ADS}[ADS]{Automated Driving System}
  \acrodef{AV}[AV]{Automated Vehicle}
  \acrodef{CAV}[CAV]{Connected and Automated Vehicle}
  \acrodef{C-ITS}[C-ITS]{Cooperative Intelligent Transport System}
  \acrodef{HRI}[HRI]{Highway Readiness Index}
  \acrodef{HD}[HD]{high-definition}
  \acrodef{HMI}[HMI]{human--machine interface}
  \acrodef{ISAD}[ISAD]{Infrastructure Support for Automated Driving}
  \acrodef{IVIM}[IVIM]{Infrastructure-to-Vehicle Information Message}
  \acrodef{LOSAD}[LOSAD]{Level of Service for Automated Driving}
  \acrodef{LOA}[LOA]{Levels of Automation}
  \acrodef{ODD}[ODD]{Operational Design Domain}
  \acrodef{OD}[OD]{Operational  Domain}
  \acrodef{RSU}[RSU]{Road Side Unit}
  \acrodef{SAE}[SAE]{Society of Automotive Engineers}
  \acrodef{V2X}[V2X]{Vehicle-to-Everything}

\title{Highway Readiness Assessment for SAE Levels of Automation and V2X Notification}

\author{\IEEEauthorblockN{Lorenzo Italiano\IEEEauthorrefmark{1}, Federico Marino\IEEEauthorrefmark{1}, Mattia Brambilla\IEEEauthorrefmark{1}, Giovanni Megna\IEEEauthorrefmark{2}, \\ Benedetto Carambia\IEEEauthorrefmark{2}, Monica Nicoli\IEEEauthorrefmark{1}}
\IEEEauthorblockA{\IEEEauthorrefmark{1}Politecnico di Milano, Milan, Italy}
\IEEEauthorblockA{\IEEEauthorrefmark{2}Movyon S.p.A., Autostrade Per l'Italia group, Florence, Italy}}


\maketitle
\begin{abstract}
While highway automation is advancing rapidly, road operators still lack practical methods to assess the readiness of their infrastructures for supporting automated driving systems. This work proposes a quantitative \ac{HRI} that maps static \ac{ODD} infrastructure conditions into measurable attributes and weights them through an expert survey to evaluate readiness across \ac{SAE} automation levels. A real corridor case study shows how \ac{HRI} scores can be computed, interpreted, and used to identify infrastructure gaps that limit higher automation. Finally, we outline how these indicators can be integrated into a standardized \ac{C-ITS} message, i.e., \ac{IVIM}, to communicate segment-level automation guidance to connected vehicles.
\end{abstract}

\acresetall


\section{Introduction}
The mobility landscape is undergoing a profound transformation driven by the urgent need to address systemic challenges related to safety, congestion, and environmental sustainability. Road accidents continues to cause over one million fatalities each year \cite{WHO_RoadTrafficInjuries_FactSheet}, while transportation remains a major contributor to global emissions \cite{IEA_Transport_CO2_2023}. In this context, \ac{AuD} technologies, embodied in \acp{CAV}, offer a promising pathway to improve safety, efficiency, and sustainability.

A key enabler of this transition is the capability of the road infrastructure to support the different \ac{LOA} defined by the \ac{SAE}. Although current standards establish clear definitions for \acp{LOA}, \acp{ODD}, and infrastructure digitalization frameworks \cite{SAE_J3016,ISO_34503_2023}, they lack a quantitative method to assess whether specific highway segments meet the operational requirements of \acp{ADS}. Consequently, road authorities remain without a structured, reproducible approach to evaluate infrastructure readiness or to identify the physical and maintenance-related factors that hinder higher \ac{LOA} levels.

The \ac{ODD} concept specifies the environmental, roadway, and traffic conditions required for the safe operation of an \ac{ADS} \cite{ShakeriODD}. However, real-world \acp{ODD} are typically qualitative, manufacturer-specific, and rarely associated with measurable thresholds, making objective infrastructure assessment particularly challenging. Concurrently, highways are progressively evolving toward smart-road paradigms~\cite{PiavaniniMagazine} by integrating \acp{RSU} with \ac{V2X}communications for \ac{C-ITS} services \cite{italianoADR,italianoAnomaly}, yet large portions of the road network still lack the digital infrastructure to support these advancements. 
\begin{figure}[t]
    \centering
    \begin{tikzpicture}
        \node[inner sep=0] (img) {\includegraphics[width=0.48\textwidth]{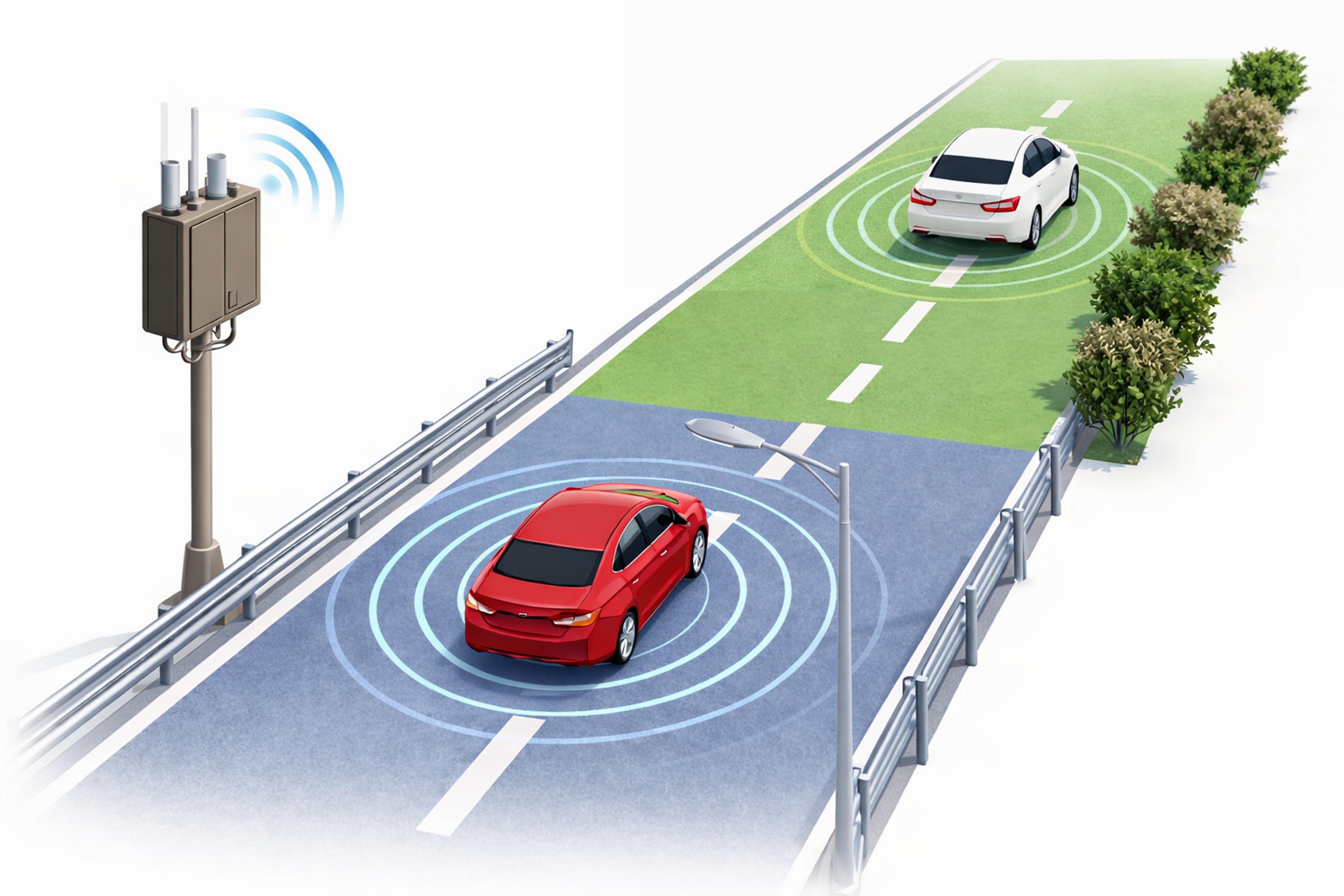}};
        \node[font=\footnotesize, fill=white] at (-3.05,2.15) {RSU};
        \draw[black, -Stealth] (-2.5,1.3)--(1.3,1.7)
            node[midway, fill=applegreen!60, inner sep=0.2mm, font=\scriptsize, rounded corners=1.5pt, align=center] {
                \begin{tcolorbox}[  colback=applegreen!10!white,
                colframe = applegreen!8!white,
                width=12mm,height=3mm,
                top=-0.9mm, left=2mm,
                arc = 0.8pt]
                IVIM
                \end{tcolorbox} \\ 
                \begin{tcolorbox}[  colback=applegreen!60,
                colframe = applegreen!60,
                width=12mm,height=3.5mm,
                top=-0.7mm, left=0mm,
                arc = 0.8pt]
                \mbox{SAE~1--2}
                \end{tcolorbox}};
        \draw[black, -Stealth] (-2.6,0.8)--(-1.2,-0.5)
            node[midway,fill=bluegray!80, inner sep=0.2mm, font=\scriptsize, rounded corners=1.5pt, align=center] {
                \begin{tcolorbox}[  colback=bluegray!10!white,
                colframe = bluegray!7!white,
                width=12mm,height=3mm,
                top=-0.9mm, left=2mm,
                arc = 0.8pt]
                IVIM
                \end{tcolorbox} \\  \begin{tcolorbox}[  colback=bluegray!80,
                colframe = bluegray!80,
                width=12mm,height=3.5mm,
                top=-0.7mm, left=0mm,
                arc = 0.8pt]
                \mbox{SAE~3--4}
                \end{tcolorbox}};
    \end{tikzpicture}
    \caption{The RSU communicates the infrastructure readiness level to CAVs via IVIM. The blue section indicates better infrastructure (SAE~3--4), while the green section is degraded by obstructive vegetation (SAE~1--2).}
    \label{fig:example_image} 
    \vspace{-5pt} 
\end{figure}
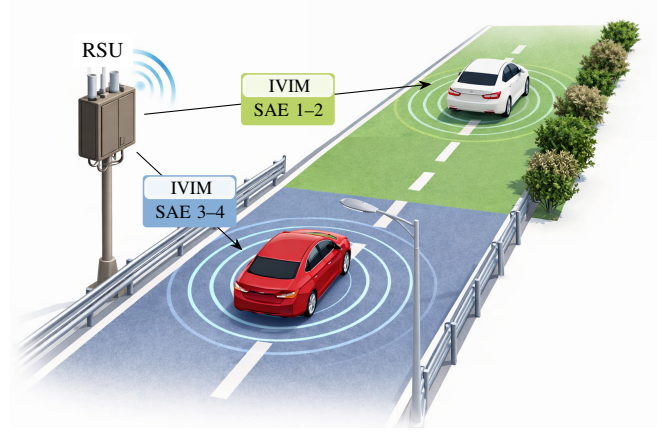

This work addresses the aforementioned gap by proposing a structured methodology aimed to quantify the highway readiness for driving automation. We introduce a new metric referred to as \ac{HRI}, which translates \ac{ODD} requirements into measurable static and semi-static\footnote{{Semi-static conditions refer to temporary but relatively stable infrastructure changes, such as roadworks or maintenance activities, and exclude dynamic factors such as traffic or weather conditions.}} road infrastructure attributes, derived and combined based on an expert survey reflecting their impact on \ac{ADS} performance. The proposed framework enables road operators to \textit{(i)} assess the capability of highway segments to support different \ac{SAE} levels, \textit{(ii)} identify limiting infrastructure factors, and \textit{(iii)} communicate recommended automation levels to vehicles through standardized \acp{IVIM}, as illustrated in Fig.~\ref{fig:example_image}. 
The resulting approach provides an infrastructure-centric and operationally interpretable framework to support evidence-based deployment of automated mobility.
\restoregeometry \noindent 
\subsection{Related Works}
Early research efforts have attempted to assess infrastructure readiness for automated mobility. 
Frameworks such as  Infrastructure Support for Automated Driving (ISAD)~\cite{INFRAMIX_ISAD_2019} and \ac{LOSAD}~\cite{PIARC_SmartRoads_Classification_2021} classify road infrastructure according to digitalization and automation support levels. 
However, ISAD mainly emphasizes connectivity and digital capabilities rather than static roadway characteristics, while \ac{LOSAD} relies on extensive fleet testing limiting scalability for large-scale network assessment. 

In parallel, significant research efforts have focused on formalizing and structuring the concept of \ac{ODD}. 
Works such as~\cite{ShakeriODD} and~\cite{HartmanODD} clarify the conceptual distinction between \ac{OD} and \ac{ODD}, particularly in the context of vehicle approval and regulatory processes. 
Taxonomy-oriented approaches~\cite{Mendiboure2023_ODD_Taxonomy} and stakeholder analyses~\cite{Mehlhorn2025_ODD_StakeholderAnalysis} further organize \ac{ODD} representations across the \ac{ADS} lifecycle, while comparative legislative studies~\cite{IvanovODD} analyze how different regulatory frameworks address \ac{ODD} specification and safety assurance. 
These contributions significantly improve conceptual clarity and alignment with government frameworks but remain primarily focused on definition, structure, and compliance.
Dynamic \ac{ODD} management enabled by \ac{V2X} communications has also been investigated in~\cite{V2XODD}, where connectivity is used to adapt the vehicle’s \ac{ODD} in real time based on traffic conditions, weather events, or temporary hazards. 
Such approaches are inherently vehicle-centric and aim at modifying operational boundaries according to transient environmental changes.

A practical example of \ac{ODD} specification is provided by the analysis of Tesla Autopilot in~\cite{Cho_Hansman_MIT_ODD_2020}, which shows that current Level~2 systems operate reliably only under constrained highway conditions and request human intervention when encountering urban complexity, adverse weather, or degraded infrastructure features. 
This illustrates how automation capability strongly depends on specific roadway characteristics, even when these constraints are not expressed through measurable infrastructure indicators.

In contrast to dynamic or vehicle-centered approaches, our work focuses on static {and semi-static} infrastructure conditions. 
Rather than adapting the \ac{ODD} in real time, we quantify how fixed roadway attributes support different \acp{LOA} and enable \acp{RSU} to communicate the appropriate \ac{SAE} levels for a given highway segment. 
Despite existing advances, a quantitative and scalable methodology that systematically maps static roadway characteristics to \ac{ADS} operational requirements is still missing. 
In particular, no current framework translates \ac{ODD} constraints into measurable infrastructure attributes that road authorities can use to evaluate automation compatibility at network scale.
To fill this gap, we introduce the \ac{HRI}, a quantitative infrastructure-centric index derived exclusively from static roadway attributes and weighted according to their impact on \ac{ADS} performance. 
Unlike digitalization-focused classifications or vehicle-dependent assessments, the proposed approach enables reproducible, large-scale evaluation of highway readiness for different \ac{SAE} levels.


\section{ODD Taxonomy and Expert Survey}
\label{sec:odd_req_survey}

We focus on static and semi-static \ac{ODD} attributes related to road infrastructure, excluding dynamic and environmental factors that are beyond the direct control of road operators. 
Furthermore, we consider two groups of \ac{LOA}: SAE~1--2 are labeled as \ac{AsD}, representing assisted driving and partial automation functions where the human driver is responsible for monitoring the driving environment, and SAE~3--4 as \ac{AuD}, corresponding to conditional and high automation where the system assumes primary control within its \ac{ODD}. 
This distinction reflects different levels of reliance on infrastructure support between supervision-based and system-dominant driving paradigms. {SAE~0 and 5 are excluded from the analysis: SAE~0 corresponds to the absence of automation and does not impose specific infrastructure requirements, while SAE~5 represents full automation, defined as the capability to operate under all environmental conditions without reliance on infrastructure support.}
The static attributes considered are grouped into four macro-categories: \emph{Road Markings \& Signage}, \emph{Road Maintenance \& Management}, \emph{Roadway Design \& Safety Features}, and \emph{Preloaded \ac{HD} Maps}, as shown in Fig.~\ref{fig:mean_impact_attributes}. 
Together, they capture the physical and structural elements that most directly influence \ac{ADS} performance.

\def\A{102.85}
\def\B{51.42}
\def\C{0}
\def\D{308.56}
\def\E{257.13}
\def\F{205.7}
\def\G{154.27}

\begin{figure}[t]
    \centering
    \begin{tikzpicture}
\begin{polaraxis}[
    width=0.8\columnwidth,
    ymin=0, ymax=5,
    ytick={1,2,3,4,5},
    yticklabel pos=right,
    yticklabel style={font=\small},
    grid=both,
    axis line style={draw=none},
    tick style={draw=none},
    xtick={\A,\B,\C,\D,\E,\F,\G},
    xticklabels={
      Professor,
      Postdoc,
      PhD\\Student,
      Technical\\Director,
      Researcher,
      R\&D\\Manager,
      Project\\Office
    },
    xticklabel style={align=center},
    legend style={
      draw=black,
      fill=white,
      font=\small,
      at={(0.82,0.78)},
      anchor=west
    },
    legend cell align=left,
]

\addplot[
    very thick,
    color=blue!70!black,
    fill=blue!40,
    fill opacity=0.18
] coordinates {
    (\A,4) (\B,4) (\C,2.8) (\D,5) (\E,4.7) (\F,3) (\G,4) (\A,4)
};
\addlegendentry{AV Exp.}

\addplot[
    very thick,
    color=orange!90!black,
    fill=orange!60,
    fill opacity=0.18
] coordinates {
    (\A,5) (\B,3) (\C,2.5) (\D,3) (\E,4.4) (\F,3.4) (\G,4) (\A,5)
};
\addlegendentry{C-ITS Exp.}

\end{polaraxis}
\end{tikzpicture}
    \caption{Survey participants' profiles by role for AV and C-ITS domains.}
    \label{fig:survey_radar}
    \vspace{-12pt}
\end{figure}
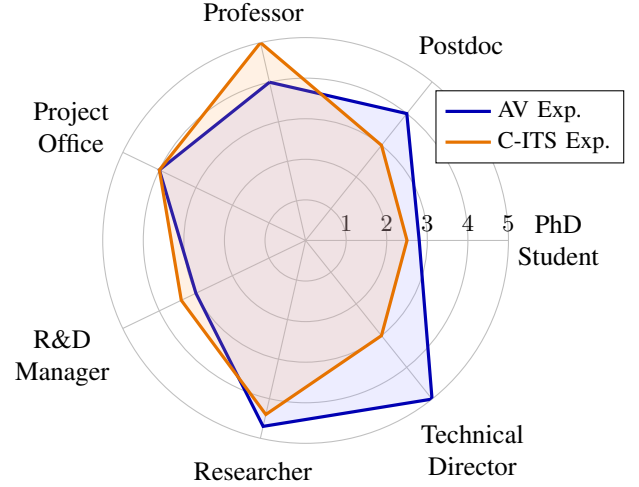

To assign meaningful weights to these attributes within the proposed \ac{HRI}, a dedicated expert survey was conducted involving 17 professionals from academia, industry, and infrastructure management. Figure~\ref{fig:survey_radar} reports their self-assessed expertise in \acp{AV} and \ac{C-ITS} on a 1--5 scale.
The survey pursued two complementary objectives: 
\textit{(i)} deriving empirical weights for each static attribute for \ac{AsD} (SAE~1--2) and \ac{AuD} (SAE~3--4), and 
\textit{(ii)} collecting expert perspectives on how digitalization and cooperative services may support or extend \ac{ADS} operational capabilities.



\pgfplotsset{compat=1.18}
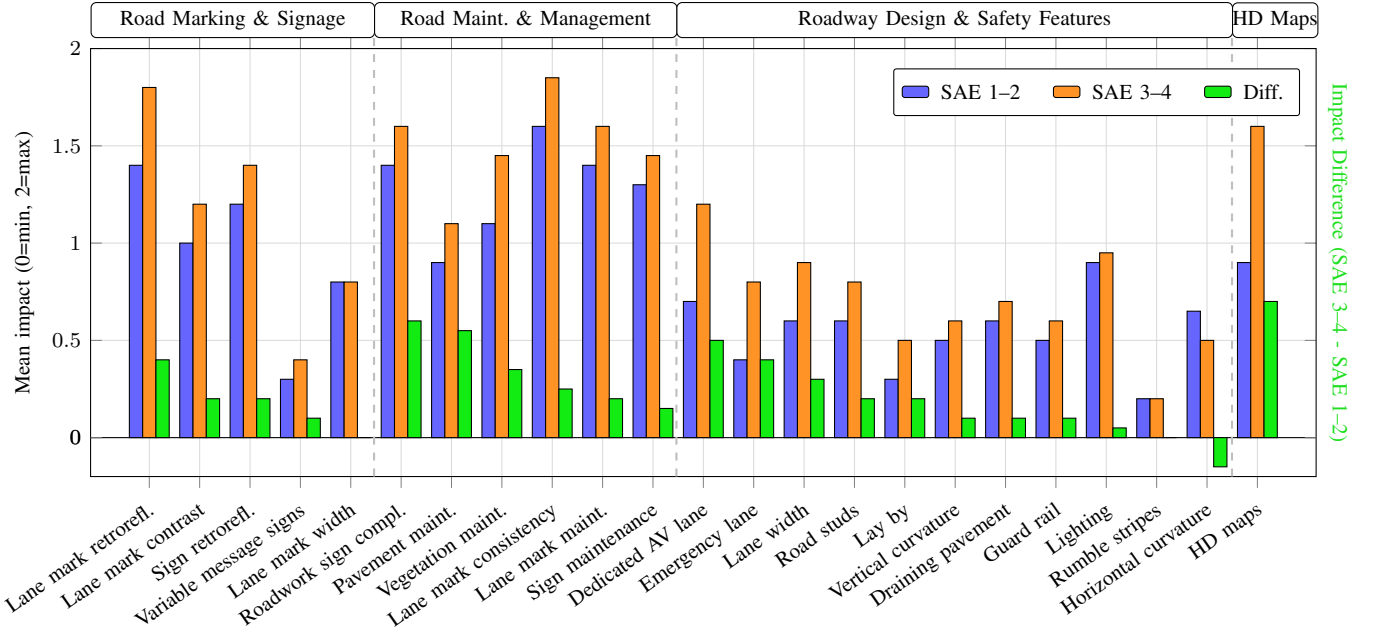
\begin{figure*}
\centering
\begin{tikzpicture}
  \begin{axis}[
    width=\textwidth,
    height=0.3\textheight,
    ybar=0pt,
    bar width=5pt,
    ymin=-0.2, ymax=2,
    ylabel={Mean impact (0=min, 2=max)},
    extra y ticks={0},
    extra y tick style={
      grid=major,
      major grid style={draw=black, line width=0.4pt},
    },
    enlarge x limits=0.053,
    xtick=data,
    xticklabel style={rotate=35, anchor=north east, font=\footnotesize},
    tick label style={font=\footnotesize},
    symbolic x coords={
      Lane mark retrorefl.,
      Lane mark contrast,
      Sign retrorefl.,
      Variable message signs,
      Lane mark width, 
      Roadwork sign compl.,
      Pavement maint.,
      Vegetation maint.,
      Lane mark consistency,
      Lane mark maint., 
      Sign maintenance,
      Dedicated AV lane,
      Emergency lane,
      Lane width,
      Road studs,
      Lay by,
      Vertical curvature,
      Draining pavement,
      Guard rail,
      Lighting,
      Rumble stripes,
      Horizontal curvature,
      HD maps,
    },
    grid=both,
    grid style={line width=.1pt, draw=gray!20},
    major grid style={line width=.2pt, draw=gray!30},
    ylabel style={font=\footnotesize},
    extra description/.append code={%
      \node[rotate=-90, anchor=north, font=\footnotesize, text=green!75!gray]
        at (rel axis cs:1.033,0.5) {Impact Difference (SAE 3--4 - SAE 1--2)};%
    },
    x tick label style={align=right},
  ]

  \addplot+[draw=black, fill=blue!60]
    coordinates {
      (Lane mark retrorefl.,1.4)
      (Lane mark width,0.8)
      (Lane mark contrast,1.0)
      (Lane mark maint.,1.4)
      (Lane mark consistency,1.6)
      (Lane width,0.6)
      (Horizontal curvature,0.65)
      (Vertical curvature,0.5)
      (Emergency lane,0.4)
      (Draining pavement,0.6)
      (Pavement maint.,0.9)
      (Lay by,0.3)
      (Rumble stripes, 0.2)
      (Sign retrorefl.,1.2)
      (Sign maintenance,1.3)
      (Variable message signs,0.3)
      (Road studs,0.6)
      (Vegetation maint.,1.1)
      (Guard rail,0.5)
      (Roadwork sign compl.,1.4)
      (Lighting,0.9)
      (Dedicated AV lane,0.7)
      (HD maps, 0.9)
    };

  \addplot+[draw=black, fill=orange!85]
    coordinates {
      (Lane mark retrorefl.,1.8)
      (Lane mark width,0.8)
      (Lane mark contrast,1.2)
      (Lane mark maint.,1.6)
      (Lane mark consistency,1.85)
      (Lane width,0.9)
      (Horizontal curvature,0.5)
      (Vertical curvature,0.6)
      (Emergency lane,0.8)
      (Draining pavement,0.7)
      (Pavement maint.,1.1)
      (Lay by,0.5)
      (Rumble stripes, 0.2)
      (Sign retrorefl.,1.4)
      (Sign maintenance,1.45)
      (Variable message signs,0.4)
      (Road studs,0.8)
      (Vegetation maint.,1.45)
      (Guard rail,0.6)
      (Roadwork sign compl.,1.6)
      (Lighting,0.95)
      (Dedicated AV lane,1.2)
      (HD maps, 1.6)
    };

  \addplot+[draw=black, fill=green!85!gray]
    coordinates {
      (Lane mark retrorefl.,0.4)
      (Lane mark width,0)
      (Lane mark contrast,0.2)
      (Lane mark maint.,0.2)
      (Lane mark consistency,0.25)
      (Lane width,0.3)
      (Horizontal curvature,-0.15)
      (Vertical curvature,0.1)
      (Emergency lane,0.4)
      (Draining pavement,0.1)
      (Pavement maint.,0.55)
      (Lay by,0.2)
      (Rumble stripes, 0)
      (Sign retrorefl.,0.2)
      (Sign maintenance,0.15)
      (Variable message signs,0.1)
      (Road studs,0.2)
      (Vegetation maint.,0.35)
      (Guard rail,0.1)
      (Roadwork sign compl.,0.6)
      (Lighting,0.05)
      (Dedicated AV lane,0.5)
      (HD maps, 0.7)
    };

  \end{axis}

  \node[anchor=south west, draw=black, rounded corners=2pt, fill=white, font=\footnotesize, inner xsep=0pt, inner ysep=3pt, minimum width = 3.75cm, align=center]
    at (0,5.8) {\, Road Marking \& Signage \,};
  \node[anchor=south west, draw=black, rounded corners=2pt, fill=white, font=\footnotesize, inner xsep=0pt, inner ysep=3pt, minimum width = 4.0cm, align=center]
    at (3.75,5.8) {Road Maint. \& Management};
  \node[anchor=south west, draw=black, rounded corners=2pt, fill=white, font=\footnotesize, inner xsep=0pt, inner ysep=3pt, minimum width = 7.35cm, align=center]
    at (7.75,5.8) {Roadway Design \& Safety Features};
  \node[anchor=south west, draw=black, rounded corners=2pt, fill=white, font=\footnotesize, inner xsep=0pt, inner ysep=3pt]
    at (15.1,5.8) {HD Maps};
  \draw[lightgray, dashed, thick] (3.75,5.8) -- (3.75,0);
  \draw[lightgray, dashed, thick] (7.75,5.8) -- (7.75,0);
  \draw[lightgray, dashed, thick] (15.1,5.8) -- (15.1,0);

    \node[
        anchor=north east,
        fill=white,
        draw=black,
        rounded corners=1pt,
        font=\footnotesize,
        inner sep=3pt
    ] at (16,5.4) {
      \begin{tikzpicture}[baseline]
        \draw[fill=blue!60, draw=black] (0,0) rectangle (0.35,0.15);
        \node[anchor=west] at (0.4,0.07) {SAE 1--2};
        
        \draw[fill=orange!85, draw=black] (2,0) rectangle (2.35,0.15);
        \node[anchor=west] at (2.4,0.07) {SAE 3--4};

        \draw[fill=green!85!gray, draw=black] (4,0) rectangle (4.35,0.15);
        \node[anchor=west] at (4.4,0.07) {Diff.};
      \end{tikzpicture}
    };

\end{tikzpicture}
\caption{Mean impact of each infrastructure attribute on \ac{ADS} operation for SAE~1--2 (blue) and SAE~3--4 (orange){, and their mean impact difference for \ac{SAE}-level groups (green)}. Above the chart are the macro-categories used to group the attributes in the questionnaire.}
\label{fig:mean_impact_attributes}
\vspace{-10pt}
\end{figure*}
The questionnaire consisted of two sections. 
The first addressed static roadway attributes derived from established \ac{ODD} taxonomies. For each attribute, respondents assessed how its absence or degradation would affect \ac{ADS} operation at SAE~1--2 and SAE~3--4 using a three-level scale, with scores of 0 (minimal impact), 1 (moderate impact), and 2 (strong impact). 
The second section investigated the role of connectivity, \ac{C-ITS} services, and \ac{HD} maps, including whether \textit{Day~1}, \textit{Day~2}, and \textit{Day~3} cooperative services could realistically extend the \ac{ODD} of an \ac{AV}. Participants were selected based on their direct involvement in \ac{AV} and smart-road technologies to ensure technically informed responses.

The quantitative results, summarized in Fig.~\ref{fig:mean_impact_attributes}, indicate that for SAE~1--2 the most influential attributes relate to lane visibility and maintenance, including retroreflectivity, marking consistency, and marking upkeep. Sign visibility, vegetation control, and pavement quality also show significant impact, confirming that \ac{AsD} primarily relies on clear and continuous visual cues.
For SAE~3--4, experts assign systematically higher importance to nearly all attributes. The most pronounced increases{, as highlighted by the impact difference analysis (in green),} are observed for \ac{HD} maps, emergency lanes, and dedicated \ac{AV} lanes, reflecting the need for higher automation levels to depend on stable roadway geometry, reliable fallback areas, and redundancy provided by detailed digital maps. When aggregated at macro-category level, \emph{Preloaded \ac{HD} Maps} and \emph{Road Maintenance \& Management} exhibit the largest increase in relevance from SAE~1--2 to SAE~3--4.

Regarding digitalization, experts broadly agree that connectivity will play an increasingly important role in future automation. \ac{C-ITS} services are considered particularly valuable for extending perception beyond onboard sensor limits, especially for SAE~3--4. As shown in Fig.~\ref{fig:cits_regions}, European respondents tend to attribute higher relevance to real-time \ac{HD} maps and advanced cooperative services (\textit{Day~2/3}), whereas U.S. respondents adopt a more vehicle-centric perspective, assigning relatively greater importance to basic awareness-oriented \textit{Day~1} services.

\pgfplotsset{compat=1.18}
\begin{figure}[t]
    \centering
    \begin{tikzpicture}
        \begin{axis}[
            width=0.49\textwidth,
            height=0.26\textheight,
            ybar=0pt,
            bar width=24pt,
            ymin=0, ymax=2.0,
            ylabel={Mean impact},
            symbolic x coords={Europe, USA},
            enlarge x limits=0.5,
            xtick=data,
            tick label style={font=\footnotesize},
            ylabel style={font=\footnotesize},
            x tick label style={font=\footnotesize},
            grid=both,
            grid style={line width=.1pt, draw=gray!20},
            major grid style={line width=.2pt, draw=gray!30},
        ]
            \addplot+[draw=black, fill=blue!70] coordinates {(Europe,1.15) (USA,1.33)};
            \addplot+[draw=black, fill=orange!85] coordinates {(Europe,1.65) (USA,1.00)};
            \addplot+[draw=black, fill=green!85!gray] coordinates {(Europe,1.70) (USA,0.33)};

        \end{axis}
    \node[
        anchor=north east,
        fill=white,
        draw=black,
        rounded corners=1pt,
        font=\footnotesize,
        inner sep=2.5pt
    ] at (7.03,4.6) {
      \begin{tikzpicture}[baseline]
        \draw[fill=green!85!gray, draw=black] (0,0) rectangle (0.35,0.15);
        \node[anchor=west] at (0.4,0.07) {C-ITS \textit{Day 3}};
        
        \draw[fill=orange!85, draw=black] (0,0.4) rectangle (0.35,0.55);
        \node[anchor=west] at (0.4,0.47) {C-ITS \textit{Day 2}};
        
        \draw[fill=blue!60, draw=black] (0,0.8) rectangle (0.35,0.95);
        \node[anchor=west] at (0.4,0.87) {C-ITS \textit{Day 1}};
      \end{tikzpicture}};
    \end{tikzpicture}
    \caption{Perceived impact of \ac{C-ITS} \textit{Day~1--3} services on extending \ac{ADS} capabilities, by region.}
    \label{fig:cits_regions}
    \vspace{-5pt}
\end{figure}
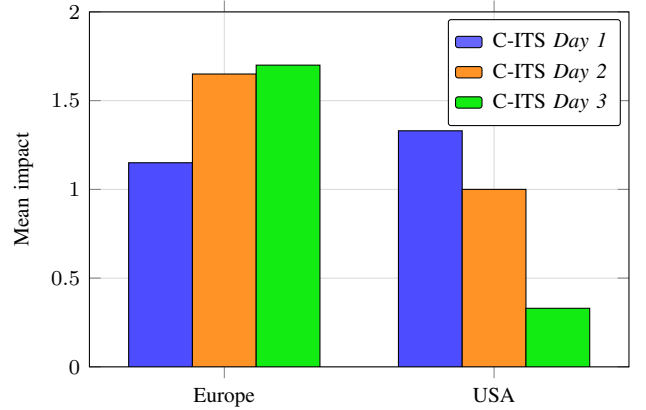


\section{Framework and Methodology}
\label{sec:methodology}

The proposed \ac{HRI} translates static infrastructure characteristics into a quantitative and operational measure of automation compatibility. 
The framework is structured into three consecutive stages, progressively converting qualitative infrastructure conditions into actionable \ac{SAE}-level guidance.

\subsection{Stage 1 -- Attribute Operationalization and Weighting}

In the first stage, each selected \ac{ODD} attribute is operationalized through measurable criteria derived from national highway design and safety standards. 
Given a road segment of arbitrary length (we consider 100~m), an observed value $v_i \in \{0,1,2\}$ is assigned, representing increasing levels of infrastructure adequacy. 
This step ensures objectivity and reproducibility in corridor-level assessments.

To reflect the different sensitivity of automation levels to infrastructure quality, each attribute is associated with two distinct weights $w_i^{(\ell)}$: one for \ac{AsD} (SAE~1--2, $\ell=1$) and one for \ac{AuD} (SAE~3--4, $\ell=2$). 
These weights, derived from the expert survey and visible in Fig.~\ref{fig:mean_impact_attributes}, transform qualitative domain knowledge into quantitative parameters, enabling differentiation between supervision-based and system-dominant automation paradigms.

\subsection{Stage 2 -- Segment-Level Readiness Model}

In the second stage, operationalized attributes $v_i^{}$ and corresponding weights $w_i^{(\ell)}$ are integrated into the \ac{HRI} scoring model. 
For each segment $i$ and automation level $\ell \in \{1,2\}$, the readiness score is computed as:

\begin{equation}
S^{(\ell)} = 
\frac{\sum_i w_i^{(\ell)} \cdot v_i}
{\sum_i w_i^{(\ell)} \cdot v_{\text{max}}}
\times 100 .    
\end{equation}

The normalization term ensures comparability across segments, while the weighting mechanism captures the relative importance of each infrastructure element. 
The resulting score expresses, in percentage form, how closely the existing roadway conditions align with the operational needs of \acp{ADS} at the considered automation level.

\subsection{Stage 3 -- Infrastructure Guidance and IVIM Integration}

The final stage converts the readiness score into operational guidance. 
From an infrastructure management perspective, the \ac{HRI} identifies high-impact attributes with insufficient adequacy, enabling prioritized and cost-effective improvement planning.

\begin{figure}
\centering
    \begin{tikzpicture}
        \node (IVIM2) [fill=lightgray!30, rounded corners = 1, anchor=north, text width = 0.97\columnwidth, minimum height = 2.3cm] at (0,-0.2-3.2){};
        \node[align=right] at (-3.5,-3.65) {\small IVIM};

        \node (IH) [fill=darkcerulean!80, anchor=north, text width=1.7cm, align = center, text depth = 1.2cm, rounded corners = 1, text=white, font=\scriptsize\linespread{0.8}\selectfont] at (-3,-0.7-3.2) {Header};
        \node (MC) [fill=darkcerulean!80, anchor=north, align=center, text width=1.7cm, text depth = 1.2cm, rounded corners = 1, text = white, font=\scriptsize\linespread{0.8}\selectfont] at (-1,-0.7-3.2) {Management Container};
        \node (GLC) [fill=darkcerulean!80, anchor=north, align=center, text width=1.7cm, text depth = 1.2cm, rounded corners = 1, text = white, font=\scriptsize\linespread{0.8}\selectfont] at (1,-0.7-3.2) {Geographic Location Container \textit{(optional)}};
        \node (DGC) [fill= apricot, text=darkcerulean, anchor=north, align=center, text width = 1.7cm, text depth = 1.2cm, rounded corners = 1, font=\scriptsize\linespread{0.8}\selectfont] at (3,-0.7-3.2) {Automated Vehicle Container \textit{(optional)}};
    \end{tikzpicture}
    \caption{IVIM container structure.}
    \label{fig:IVIM}
\end{figure}
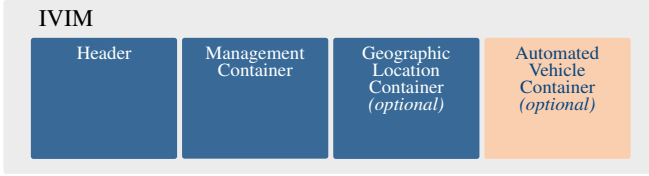

From a cooperative mobility perspective, readiness outputs are embedded into the standardized \ac{IVIM}, presented in Fig.~\ref{fig:IVIM}. This message is used to transmit information from road infrastructure to vehicles about traffic, roadway status, and weather conditions~\cite{IVIM}. The objective is to anticipate conditions that could trigger an autonomous driving disengagement, thereby requiring greater attention from the human supervisor or a takeover.
The readiness score, normalized in the range [0,100], is classified into three interpretative categories obtained through uniform partitioning of the interval: \textit{Unlikely to be suitable} (0–33\%), \textit{May be suitable} (33–66\%), and \textit{Highly likely to be suitable} (66–100\%). The 66\% threshold therefore corresponds to the lower bound of the highest readiness class and provides a neutral, data-independent criterion for conservative \ac{SAE}-level recommendations.

{Only SAE-levels whose readiness score are above the 66\% threshold are included in the \texttt{allowedSaeAutomationLevels} field of the \texttt{Automated Vehicle Container}. When multiple levels satisfy this condition, all are communicated, with the highest level representing the maximum infrastructure-supported automation capability.}
This ensures interoperable and infrastructure-driven dissemination of recommended automation levels to vehicles via \ac{C-ITS}.
Alternative thresholds could be calibrated using empirical performance data in future studies; however, the current choice ensures transparency and methodological consistency.

\begin{figure}[t]
    \centering
    \begin{tikzpicture}
        \node[inner sep=0] (img) {\includegraphics[width=0.95\columnwidth]{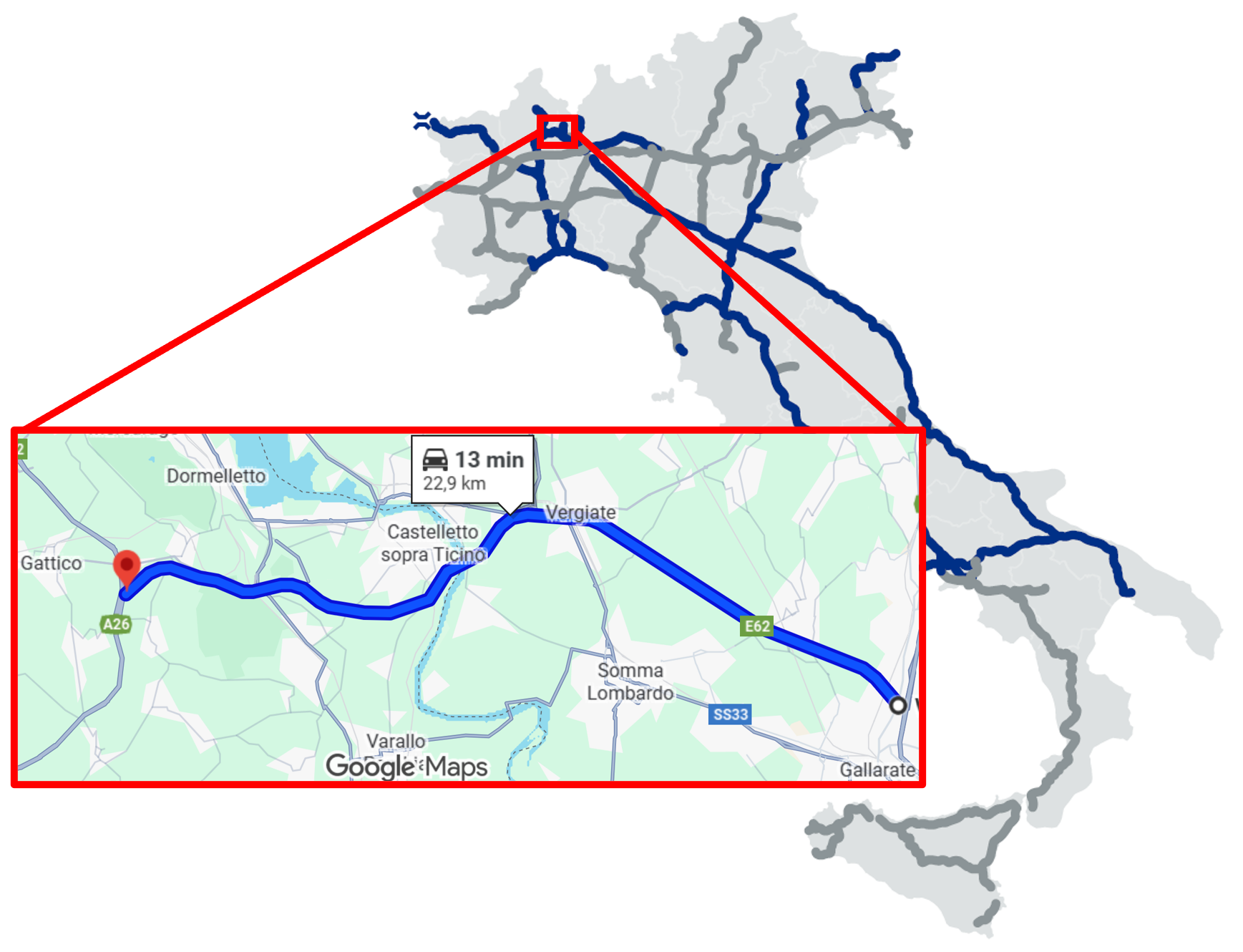}};
        \node[fill=white, inner sep=1.35pt, font=\fontsize{4}{6}\selectfont] at (-1.14,-0.06) {24\,km\,\,};
    \end{tikzpicture}
    \caption{Gallarate--Gattico corridor (D08-A8/A26) and analyzed ASPI highway section.}
    \label{fig:gallarate_section}
\end{figure}

\section{Case Study}
\label{sec:case_study}

The proposed readiness assessment methodology is applied to a real highway corridor to evaluate its practical behavior under different infrastructure configurations. 
The selected section, shown in Fig.~\ref{fig:gallarate_section}, corresponds to the D08 Gallarate--Gattico branch, a 24~km corridor of the A8/A26 system in Northern Italy, managed by \ac{ASPI}.

A dedicated data collection campaign was conducted to operationalize all static \ac{ODD} attributes defined in Sec.~\ref{sec:odd_req_survey}. 
For each 100~m segment, observable values were extracted from infrastructure inspection records, design documentation, maintenance reports, and on-site verification. 
This process ensured that all readiness scores reflect measurable, real-world conditions rather than simulated assumptions.
The resulting \ac{SAE}-level readiness scores are summarized in Fig.~\ref{fig:readiness_comparison}.

\subsection{Baseline Conditions}

\begin{figure}[t]
    \centering
    \subfloat[Baseline conditions (SAE~1--2 and SAE~3--4). \label{fig:baseline_profile}]{
        \centering
        \begin{tikzpicture}
\begin{axis}[
    width=\columnwidth,
    height=0.2\textheight,
    xmin=0, xmax=24,
    ymin=40, ymax=100,
    xlabel={[km]},
    ylabel={Score [\%]},
    xtick={0,2,4,6,8,10,12,14,16,18,20,22,24},
    ytick={40,50,60,70,80,90,100},
    grid=major,
    grid style={dashed,gray!30},
    tick label style={font=\footnotesize},
    label style={font=\footnotesize},
    minor tick num=0,
    smooth,
    mark options={solid,fill=white},
    every axis plot/.append style={line width=0.9pt},
    legend style={
        at={(0.98,0.02)},
        anchor=south east,
        font=\footnotesize
    }
]

\addplot[
    blue!70!black,
] table[
    y expr=\thisrowno{0},
    x expr=\coordindex*0.1,
    row sep=\\
] {
78\\
78\\
80\\
80\\
80\\
78\\
78\\
78\\
75\\
75\\
78\\
76\\
78\\
78\\
78\\
75\\
75\\
75\\
75\\
78\\
78\\
76\\
73\\
73\\
73\\
73\\
73\\
73\\
76\\
78\\
78\\
78\\
78\\
78\\
78\\
78\\
78\\
78\\
81\\
84\\
78\\
81\\
81\\
78\\
78\\
78\\
75\\
75\\
78\\
78\\
78\\
78\\
78\\
78\\
78\\
78\\
78\\
78\\
78\\
81\\
84\\
84\\
84\\
81\\
81\\
78\\
78\\
78\\
78\\
75\\
75\\
78\\
78\\
78\\
78\\
78\\
78\\
80\\
80\\
80\\
80\\
80\\
80\\
80\\
80\\
80\\
80\\
80\\
80\\
80\\
80\\
80\\
80\\
80\\
80\\
80\\
80\\
80\\
80\\
80\\
80\\
80\\
80\\
80\\
80\\
80\\
81\\
81\\
81\\
81\\
81\\
81\\
81\\
81\\
81\\
86\\
86\\
81\\
84\\
84\\
81\\
77\\
77\\
76\\
77\\
82\\
82\\
82\\
82\\
82\\
82\\
82\\
77\\
77\\
77\\
77\\
82\\
82\\
82\\
82\\
82\\
82\\
82\\
79\\
77\\
77\\
82\\
82\\
82\\
82\\
82\\
82\\
81\\
81\\
81\\
81\\
81\\
81\\
81\\
81\\
81\\
81\\
81\\
81\\
81\\
81\\
78\\
78\\
78\\
78\\
81\\
81\\
81\\
81\\
83\\
86\\
86\\
84\\
81\\
81\\
81\\
81\\
81\\
79\\
79\\
81\\
81\\
81\\
81\\
78\\
78\\
78\\
79\\
79\\
79\\
79\\
81\\
81\\
81\\
81\\
81\\
81\\
81\\
81\\
81\\
81\\
81\\
81\\
81\\
81\\
81\\
86\\
83\\
83\\
78\\
78\\
80\\
80\\
80\\
80\\
80\\
80\\
80\\
80\\
80\\
80\\
80\\
80\\
80\\
80\\
80\\
80\\
80\\
80\\
80\\
80\\
80\\
80\\
85\\
82\\
82\\
82\\
};
\addplot[
    green!60!black,
] table[
    y expr=\thisrowno{0},
    x expr=\coordindex*0.1
] {
    72	
    72	
    74	
    74	
    74	
    75	
    75	
    75	
    73	
    73	
    75	
    72	
    75	
    75	
    75	
    73	
    73	
    73	
    73	
    75	
    75	
    73	
    66	
    66	
    66	
    66	
    66	
    66	
    73	
    75	
    75	
    75	
    75	
    75	
    75	
    75	
    75	
    75	
    76	
    80	
    75	
    76	
    76	
    75	
    75	
    75	
    73	
    73
    75	
    75	
    75	
    75	
    75	
    75	
    75	
    75	
    75	
    75	
    75	
    76	
    80	
    80	
    80	
    76	
    76	
    75	
    75	
    75	
    75	
    73	
    73	
    75	
    75	
    75	
    75	
    75	
    75	
    77	
    77	
    77	
    77	
    77	
    77	
    77	
    77	
    77	
    77	
    77	
    77	
    77	
    77	
    77	
    77	
    77	
    77	
    77	
    77	
    77	
    77	
    77	
    77	
    77	
    77	
    77	
    77	
    77	
    78	
    78	
    78	
    78	
    78	
    78	
    78	
    78	
    78	
    82	
    82	
    78	
    79	
    79	
    78	
    73	
    73	
    72	
    73	
    77	
    77	
    77	
    77	
    77	
    77	
    77	
    73	
    73	
    73	
    73	
    77	
    77	
    77	
    77	
    77	
    77	
    77	
    77	
    73	
    73	
    77	
    77	
    77	
    77	
    77	
    77	
    78	
    78	
    78	
    78	
    78	
    78	
    79	
    79	
    79	
    79	
    79	
    79	
    79	
    79	
    76	
    76	
    76	
    76	
    79	
    79	
    79	
    78	
    80	
    82	
    82	
    79	
    79	
    79	
    79	
    79	
    79	
    75	
    75	
    79	
    79	
    79	
    79	
    76	
    76	
    76	
    75	
    75	
    75	
    75	
    79
    79	
    79	
    79	
    79	
    79	
    79	
    79	
    78	
    78	
    78	
    78	
    78	
    78	
    79	
    83	
    78	
    78	
    74	
    74	
    78	
    78	
    78	
    78	
    78	
    78	
    78	
    78	
    78	
    78	
    78	
    78	
    78	
    78	
    78	
    78	
    78	
    78	
    78	
    77	
    77	
    77	
    82	
    76	
    76	
    76
};


\addplot[red!60!black, dashed, thick] coordinates {(0,66) (24,66)};
\legend{SAE 1--2, SAE 3--4}
\end{axis}
\end{tikzpicture}}
    \hfill
    \subfloat[Roadworks with loss of consistency.\label{fig:RW_loss}]{
        \centering
        \begin{tikzpicture}
\begin{axis}[
    width=\columnwidth,
    height=0.2\textheight,
    xmin=0, xmax=24,
    ymin=40, ymax=100,
    xlabel={[km]},
    ylabel={Score [\%]},
    xtick={0,2,4,6,8,10,12,14,16,18,20,22,24},
    ytick={40,50,60,70,80,90,100},
    grid=major,
    grid style={dashed,gray!30},
    tick label style={font=\footnotesize},
    label style={font=\footnotesize},
    minor tick num=0,
    smooth,
    mark options={solid,fill=white},
    every axis plot/.append style={line width=0.9pt},
]

\addplot[
    blue!70!black,
] table[
    y expr=\thisrowno{0},
    x expr=\coordindex*0.1,
    row sep=\\
] {
78\\
78\\
80\\
80\\
80\\
78\\
78\\
78\\
75\\
75\\
78\\
76\\
78\\
78\\
78\\
75\\
75\\
75\\
75\\
78\\
78\\
76\\
73\\
73\\
73\\
73\\
73\\
73\\
76\\
78\\
78\\
78\\
78\\
78\\
78\\
78\\
78\\
78\\
81\\
84\\
78\\
81\\
81\\
78\\
78\\
78\\
75\\
75\\
78\\
78\\
78\\
78\\
78\\
78\\
78\\
78\\
78\\
78\\
78\\
81\\
84\\
84\\
84\\
81\\
81\\
78\\
78\\
78\\
78\\
75\\
75\\
78\\
78\\
78\\
78\\
78\\
78\\
80\\
80\\
80\\
80\\
80\\
80\\
80\\
80\\
80\\
80\\
80\\
80\\
80\\
80\\
80\\
80\\
80\\
80\\
80\\
80\\
80\\
80\\
80\\
80\\
80\\
80\\
80\\
80\\
80\\
81\\
81\\
81\\
81\\
50\\
50\\
50\\
50\\
50\\
55\\
55\\
50\\
55\\
55\\
50\\
50\\
50\\
49\\
50\\
69\\
69\\
69\\
69\\
69\\
69\\
69\\
50\\
50\\
50\\
50\\
69\\
69\\
69\\
69\\
69\\
69\\
69\\
50\\
50\\
50\\
69\\
69\\
69\\
69\\
69\\
69\\
50\\
50\\
50\\
50\\
50\\
50\\
50\\
50\\
50\\
50\\
50\\
50\\
50\\
50\\
48\\
48\\
48\\
48\\
50\\
50\\
81\\
81\\
83\\
86\\
86\\
84\\
81\\
81\\
81\\
81\\
81\\
79\\
79\\
81\\
81\\
81\\
81\\
78\\
78\\
78\\
79\\
79\\
79\\
79\\
81\\
81\\
81\\
81\\
81\\
81\\
81\\
81\\
81\\
81\\
81\\
81\\
81\\
81\\
81\\
86\\
83\\
83\\
78\\
78\\
80\\
80\\
80\\
80\\
80\\
80\\
80\\
80\\
80\\
80\\
80\\
80\\
80\\
80\\
80\\
80\\
80\\
80\\
80\\
80\\
80\\
80\\
85\\
82\\
82\\
82\\
};
\addplot[
    green!60!black
] table[
    y expr=\thisrowno{0},
    x expr=\coordindex*0.1,
    row sep=\\
] {
72\\
72\\
74\\
74\\
74\\
75\\
75\\
75\\
73\\
73\\
75\\
72\\
75\\
75\\
75\\
73\\
73\\
73\\
73\\
75\\
75\\
73\\
66\\
66\\
66\\
66\\
66\\
66\\
73\\
75\\
75\\
75\\
75\\
75\\
75\\
75\\
75\\
75\\
76\\
80\\
75\\
76\\
76\\
75\\
75\\
75\\
73\\
73\\
75\\
75\\
75\\
75\\
75\\
75\\
75\\
75\\
75\\
75\\
75\\
76\\
80\\
80\\
80\\
76\\
76\\
75\\
75\\
75\\
75\\
73\\
73\\
75\\
75\\
75\\
75\\
75\\
75\\
77\\
77\\
77\\
77\\
77\\
77\\
77\\
77\\
77\\
77\\
77\\
77\\
77\\
77\\
77\\
77\\
77\\
77\\
77\\
77\\
77\\
77\\
77\\
77\\
77\\
77\\
77\\
77\\
77\\
78\\
78\\
78\\
78\\
49\\
49\\
49\\
49\\
49\\
52\\
52\\
49\\
52\\
52\\
49\\
49\\
49\\
47\\
49\\
66\\
66\\
66\\
66\\
66\\
66\\
66\\
49\\
49\\
49\\
49\\
66\\
66\\
66\\
66\\
66\\
66\\
66\\
49\\
49\\
49\\
66\\
66\\
66\\
66\\
66\\
66\\
49\\
49\\
49\\
49\\
49\\
49\\
49\\
49\\
49\\
49\\
49\\
49\\
49\\
49\\
46\\
46\\
46\\
46\\
49\\
49\\
79\\
78\\
80\\
82\\
82\\
79\\
79\\
79\\
79\\
79\\
79\\
75\\
75\\
79\\
79\\
79\\
79\\
76\\
76\\
76\\
75\\
75\\
75\\
75\\
79\\
79\\
79\\
79\\
79\\
79\\
79\\
79\\
78\\
78\\
78\\
78\\
78\\
78\\
79\\
83\\
78\\
78\\
74\\
74\\
78\\
78\\
78\\
78\\
78\\
78\\
78\\
78\\
78\\
78\\
78\\
78\\
78\\
78\\
78\\
78\\
78\\
78\\
78\\
77\\
77\\
77\\
82\\
76\\
76\\
76\\
};

\addplot[red!60!black, dashed, thick] coordinates {(0,66) (24,66)};
\end{axis}
\end{tikzpicture}}
    \hfill
    \subfloat[Poor maintenance and missing safety features.\label{fig:CS_Maintenance_poor}]{
        \begin{tikzpicture}
\begin{axis}[
    width=\columnwidth,
    height=0.2\textheight,
    xmin=0, xmax=24,
    ymin=40, ymax=100,
    xlabel={[km]},
    ylabel={Score [\%]},
    xtick={0,2,4,6,8,10,12,14,16,18,20,22,24},
    ytick={40,50,60,70,80,90,100},
    grid=major,
    grid style={dashed,gray!30},
    tick label style={font=\footnotesize},
    label style={font=\footnotesize},
    minor tick num=0,
    smooth,
    mark options={solid,fill=white},
    every axis plot/.append style={line width=0.9pt},
]

\addplot[
    blue!70!black,
] table[
    y expr=\thisrowno{0},
    x expr=\coordindex*0.1,
    row sep=\\
] {78\\
78\\
80\\
80\\
80\\
78\\
78\\
78\\
75\\
75\\
78\\
76\\
78\\
78\\
78\\
75\\
75\\
75\\
75\\
78\\
78\\
76\\
73\\
73\\
73\\
43\\
43\\
43\\
52\\
54\\
54\\
54\\
54\\
54\\
54\\
54\\
54\\
54\\
52\\
54\\
54\\
52\\
52\\
54\\
54\\
54\\
51\\
51\\
54\\
54\\
54\\
54\\
54\\
54\\
54\\
54\\
54\\
54\\
54\\
52\\
54\\
54\\
54\\
52\\
52\\
54\\
54\\
54\\
54\\
51\\
51\\
54\\
54\\
54\\
54\\
54\\
54\\
56\\
56\\
56\\
56\\
56\\
56\\
56\\
56\\
56\\
56\\
56\\
56\\
56\\
56\\
56\\
56\\
56\\
56\\
56\\
56\\
56\\
56\\
56\\
56\\
56\\
56\\
56\\
56\\
56\\
57\\
57\\
57\\
57\\
57\\
57\\
57\\
57\\
57\\
57\\
57\\
57\\
55\\
55\\
57\\
54\\
54\\
52\\
54\\
54\\
54\\
54\\
54\\
54\\
54\\
54\\
54\\
54\\
54\\
54\\
54\\
54\\
54\\
54\\
54\\
54\\
54\\
56\\
54\\
54\\
54\\
54\\
54\\
54\\
54\\
54\\
57\\
57\\
57\\
57\\
81\\
81\\
81\\
81\\
81\\
81\\
81\\
81\\
81\\
81\\
78\\
78\\
78\\
78\\
81\\
81\\
81\\
81\\
83\\
86\\
86\\
84\\
81\\
81\\
81\\
81\\
81\\
79\\
79\\
81\\
81\\
81\\
81\\
78\\
78\\
78\\
79\\
79\\
79\\
79\\
81\\
81\\
81\\
81\\
81\\
81\\
81\\
81\\
81\\
81\\
81\\
81\\
81\\
81\\
81\\
86\\
83\\
83\\
78\\
78\\
80\\
80\\
80\\
80\\
80\\
80\\
80\\
80\\
80\\
80\\
80\\
80\\
80\\
80\\
80\\
80\\
80\\
80\\
80\\
80\\
80\\
80\\
85\\
82\\
82\\
82\\
};
\addplot[
    green!60!black
] table[
    y expr=\thisrowno{0},
    x expr=\coordindex*0.1,
    row sep=\\
] {72\\
72\\
74\\
74\\
74\\
75\\
75\\
75\\
73\\
73\\
75\\
72\\
75\\
75\\
75\\
73\\
73\\
73\\
73\\
75\\
75\\
73\\
66\\
66\\
66\\
39\\
39\\
39\\
50\\
52\\
52\\
52\\
52\\
52\\
52\\
52\\
52\\
52\\
48\\
52\\
52\\
48\\
48\\
52\\
52\\
52\\
49\\
49\\
52\\
52\\
52\\
52\\
52\\
52\\
52\\
52\\
52\\
52\\
52\\
48\\
52\\
52\\
52\\
48\\
48\\
52\\
52\\
52\\
52\\
49\\
49\\
52\\
52\\
52\\
52\\
52\\
52\\
54\\
54\\
54\\
54\\
54\\
54\\
54\\
54\\
54\\
54\\
54\\
54\\
54\\
54\\
54\\
54\\
54\\
54\\
54\\
54\\
54\\
54\\
54\\
54\\
54\\
54\\
54\\
54\\
54\\
56\\
56\\
56\\
56\\
56\\
56\\
56\\
56\\
56\\
56\\
56\\
56\\
52\\
52\\
56\\
51\\
51\\
49\\
51\\
51\\
51\\
51\\
51\\
51\\
51\\
51\\
51\\
51\\
51\\
51\\
51\\
51\\
51\\
51\\
51\\
51\\
51\\
55\\
51\\
51\\
51\\
51\\
51\\
51\\
51\\
51\\
56\\
56\\
56\\
56\\
78\\
78\\
79\\
79\\
79\\
79\\
79\\
79\\
79\\
79\\
76\\
76\\
76\\
76\\
79\\
79\\
79\\
78\\
80\\
82\\
82\\
79\\
79\\
79\\
79\\
79\\
79\\
75\\
75\\
79\\
79\\
79\\
79\\
76\\
76\\
76\\
75\\
75\\
75\\
75\\
79\\
79\\
79\\
79\\
79\\
79\\
79\\
79\\
78\\
78\\
78\\
78\\
78\\
78\\
79\\
83\\
78\\
78\\
74\\
74\\
78\\
78\\
78\\
78\\
78\\
78\\
78\\
78\\
78\\
78\\
78\\
78\\
78\\
78\\
78\\
78\\
78\\
78\\
78\\
77\\
77\\
77\\
82\\
76\\
76\\
76\\
};

\addplot[red!60!black, dashed, thick] coordinates {(0,66) (24,66)};
\end{axis}
\end{tikzpicture}}
    \caption{Readiness trends under different infrastructure conditions. The dashed red line represents the 66\% threshold separating the \textit{Highly likely to} and \textit{May be suitable}.}
    \label{fig:readiness_comparison}
\end{figure}
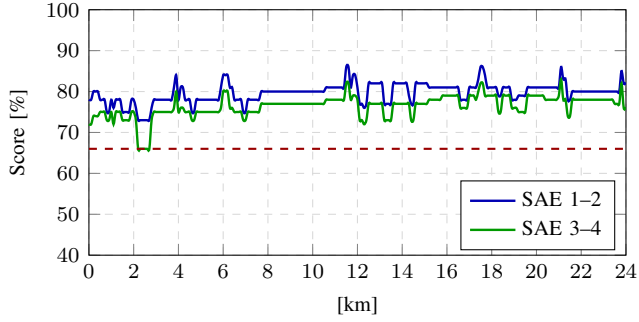
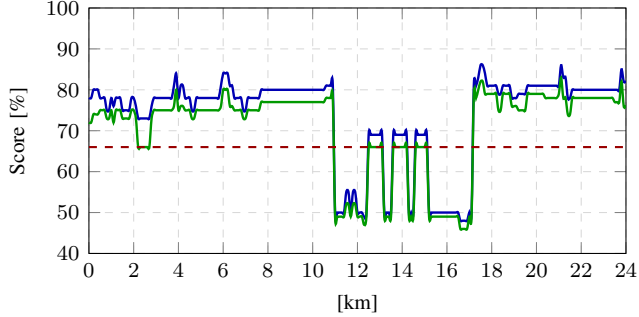
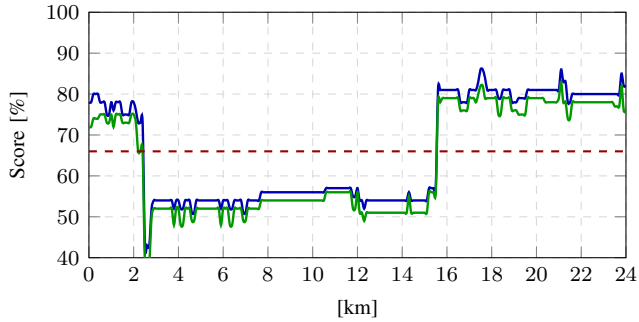

The baseline scenario evaluates the corridor under its current operational conditions. 
As shown in Fig.~\ref{fig:baseline_profile}, SAE~1--2 readiness remains consistently within the \textit{Highly likely to be suitable} range. 
SAE~3--4 scores are slightly lower, reflecting the stricter infrastructure requirements associated with higher automation levels.
Under these conditions, the generated \ac{IVIM} message would recommend SAE levels up to~4 along most of the corridor, confirming that well-maintained highway infrastructure can already support advanced automation in stable configurations.

%


\subsection{Roadworks Conditions}

To evaluate robustness under temporary{, semi-static} modifications, a second scenario models roadworks between km~11--17. 
This scenario introduces degraded temporary markings, loss of lane consistency, and poor geometric transitions near tunnel exits, which are common during roadworks.

As shown in Fig.~\ref{fig:RW_loss}, readiness decreases sharply, particularly for SAE~3--4, with multiple segments falling into the \textit{May be suitable} range. 
In this case, the corresponding \ac{IVIM} message would conservatively omit \ac{SAE}-level recommendations in the affected sections.
These results confirm that the methodology is sensitive to localized degradations and can quantify their operational impact in a segment-by-segment manner.


\subsection{Maintenance and Safety Conditions}

A third scenario investigates the impact of long-term maintenance quality and safety equipment. 
Between km~3--16, degraded markings, pavement deterioration, and missing safety devices were introduced to simulate substandard maintenance conditions.
As shown in Fig.~\ref{fig:CS_Maintenance_poor}, readiness drops significantly for both SAE~1--2 and SAE~3--4, with several segments entering the \textit{Unlikely} range. 
Under these conditions, no automation level satisfies the threshold required for \ac{IVIM} transmission.
This scenario highlights the structural dependency of automation on consistent infrastructure quality and demonstrates the ability of the \ac{HRI} to capture systemic degradation beyond temporary disturbances.


\subsection{Macro-Category Sensitivity Analysis}

To move beyond the single-corridor application and assess the structural behavior of the framework, a macro-category sensitivity analysis was conducted based solely on the theoretical weights derived from the expert survey.
Rather than using observed segment data, this analysis aggregates attribute weights into the four macro-categories introduced in Sec.~\ref{sec:odd_req_survey}, and reported in Table~\ref{tab:macro_weights}. 
This allows isolation of their relative contribution under controlled configurations, independently from the D08 corridor.

\begin{table}[t]
    \caption{Average macro-category weights from the expert survey.}
    \label{tab:macro_weights}
    \centering
    \begin{tabular}{lcc}
        \toprule
        \textbf{Macro-category} & \textbf{SAE~1--2} & \textbf{SAE~3--4} \\
        \midrule
        Road Markings \& Signage & 0.9 & 1.1 \\
        Road Maintenance \& Management & 1.3 & 1.5 \\
        Roadway Design \& Safety Features & 0.5 & 0.7 \\
        Preloaded \ac{HD} maps & 0.9 & 1.7 \\
        \bottomrule
    \end{tabular}
\end{table}

\begin{figure}[t]
    \centering
    \begin{tikzpicture}
        \node[anchor=south west, inner sep=0] (img) at (0,0)
            {\includegraphics[width=0.4\textwidth]{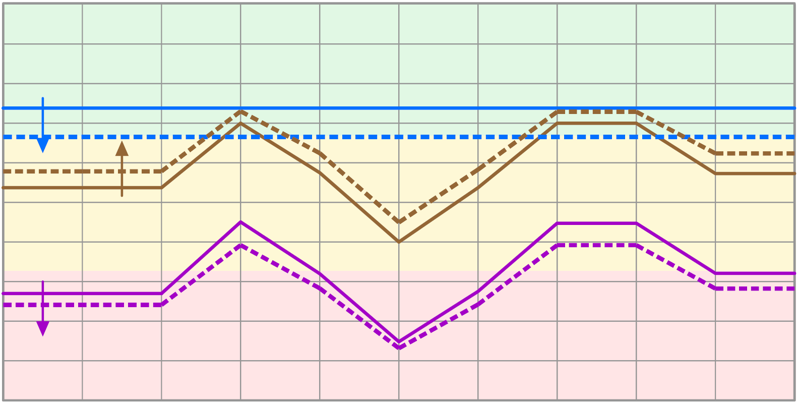}};

        \begin{scope}[x={(img.south east)},y={(img.north west)}]
            \draw[black, line width=0.4pt] (0,0) -- (1,0);
            \draw[black, line width=0.4pt] (0,0) -- (0,1);

            \foreach \i in {0,1,...,10} {
                \draw[black, line width=0.4pt] ({\i/10},0) -- ({\i/10},-0.02);
                \node[font=\scriptsize, anchor=north] at ({\i/10},-0.02) {\i};
            }
            \foreach \j in {0,10,...,100} {
                \draw[black, line width=0.4pt] (0,{\j/100}) -- (-0.02,{\j/100});
                \node[font=\scriptsize, anchor=east] at (-0.02,{\j/100}) {\j};
            }

            \node[font=\scriptsize, anchor=north] at (0.5,-0.08) {[km]};
            \node[font=\scriptsize, anchor=south, rotate=90] at (-.1,0.5) {Score [\%]};
        \end{scope}
            \node[
        anchor=north east,
        fill=white,
        draw=black,
        rounded corners=1pt,
        font=\footnotesize,
        inner sep=3pt
    ] at (7.1,3.55) {
      \begin{tikzpicture}[baseline]
        \draw[black, line width=0.4pt] (0,0.075) -- (0.4,0.075);
        \node[anchor=west] at (0.4,0.07) {SAE 1--2};
        
        \draw[black, densely dashed] (1.9,0.075) -- (2.4,0.075);
        \node[anchor=west] at (2.4,0.07) {SAE 3--4};
      \end{tikzpicture}
    };
    \end{tikzpicture}

    \caption{Macro-category infrastructure readiness trends for SAE~1--2 (solid lines) and SAE~3--4 (dashed lines), {comparing fully compliant infrastructure without \ac{HD} maps (blue), degraded infrastructure complemented with preloaded \ac{HD} maps (brown), and degraded infrastructure without \ac{HD} maps (purple)}.}
    \label{fig:macro}
    \vspace{-5pt}
\end{figure}
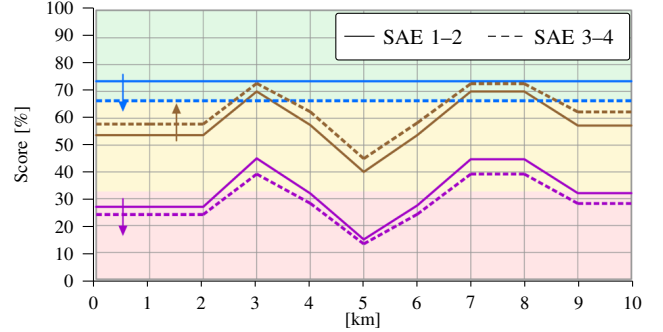

Three representative infrastructure configurations were evaluated in a progressive sequence:
 \textit{(i)} fully compliant infrastructure without \ac{HD} maps (blue lines), representing typical baseline readiness;
 \textit{(ii)} degraded infrastructure and maintenance without \ac{HD} maps (purple lines), reflecting a common real-world condition and its impact on readiness;
 \textit{(iii)} the same degraded scenario complemented with preloaded \ac{HD} maps (brown lines), highlighting how \ac{HD} maps—assigned the highest weight in Table~\ref{tab:macro_weights} significantly improve readiness, especially for SAE~3--4, and shift segments from \textit{Unlikely to be suitable} (red) to \textit{May be suitable} (yellow).

As shown in Fig.~\ref{fig:macro}, degradation significantly reduces readiness for both SAE groups, with a stronger impact on SAE~3--4 due to their higher sensitivity to infrastructure quality. 
However, the inclusion of \ac{HD} maps partially compensates for physical deficiencies, particularly for higher automation levels.
Unlike the previous corridor-based scenarios, this analysis demonstrates the intrinsic behavior of the \ac{HRI} weighting structure. 
It confirms that the methodology is not corridor-specific, but generalizable to different infrastructure networks and planning contexts.

Overall, this step validates the versatility of the framework, showing that it can operate both as a corridor-level assessment tool and as a strategic planning instrument for network-wide automation readiness evaluation.

\section{Conclusions}
\label{sec:conclusion}
{This paper presented a quantitative and infrastructure-centric methodology to assess highway readiness for automated driving through the proposed \ac{HRI} metric. 
The expert survey highlighted that lane-related attributes (particularly marking consistency, retroreflectivity, and maintenance) are key enablers across all \ac{SAE} levels, while signage quality, vegetation control, and pavement condition play a decisive role for SAE~1--2, and \ac{HD} maps are critical for SAE~3--4.
The D08 corridor case study demonstrated the framework’s applicability and sensitivity. 
Roadworks and degraded maintenance caused the largest readiness drops, whereas improved infrastructure quality and the addition of \ac{HD} maps significantly increased readiness, confirming the importance of both physical and digital layers.
\ac{SAE}-level recommendations via \ac{IVIM} consistently reflect infrastructure readiness, with only segments above the 66\% threshold enabling automation guidance.}

{Future work will extend the framework to incorporate dynamic \ac{ODD} conditions, including traffic, incidents, and environmental variability, to analyze how static \ac{HRI} varies under changing operating conditions, and will expand the expert survey at an international scale. 
Further research will investigate how \ac{C-ITS} and real-time \ac{IVIM} dissemination can support adaptive automation guidance. 
Future developments will also enrich the \ac{IVIM} content beyond SAE levels, introducing descriptive or advisory messages to provide more actionable guidance. 
Integrating static, dynamic, and digital dimensions represents the next step toward a comprehensive assessment of infrastructure readiness for \ac{AuD}, while maintaining minimum \ac{HRI} levels over time could support proactive maintenance planning.}

\section*{Acknowledgment}
{This work was funded by the Italian Ministry of University
and Research (MUR) Decree n. 352—09/04/2022 under the
National Recovery and Resilience Plan (NRRP) and European
Union (EU) NextGenerationEU project and by Autostrade per
l’Italia (ASPI). The authors acknowledge the use of ChatGPT, accessed through the Politecnico di Milano ChatGPT Edu license, for refining Fig.~\ref{fig:example_image}.}


\bibliographystyle{IEEEtran}
\bibliography{IEEEabrv, bibliography}

\end{document}